\documentclass[12pt, draftclsnofoot, onecolumn]{IEEEtran}
\usepackage{graphicx}

\usepackage[caption=false,textfont=sf]{subfig}
\usepackage{changes}
\usepackage{amsmath}

\usepackage{multicol}
\usepackage{amssymb}
\usepackage{subfig}
\usepackage{lineno}
\usepackage{makecell}
\usepackage{indentfirst}
\usepackage{bm}
\usepackage{color}
\usepackage{cite}
\usepackage{epstopdf}
\usepackage[noend]{algpseudocode}
\usepackage{algorithmicx,algorithm}
\usepackage{amssymb}
\usepackage{xcolor}
\usepackage{tikz}
\usepackage{mathtools}
\usepackage{amsthm}
\usepackage{mathrsfs}
\usepackage{verbatim}
\usepackage{epstopdf}

\newcommand{\HH}{\mathrm{H}}

\ifCLASSINFOpdf
\else
\fi
\usepackage{amsmath}
\begin{document}
	
	\title{Mutual Information-Based Integrated Sensing and Communications: A WMMSE Framework}

	\author{Yizhou Peng, Songjie Yang, Wanting Lyu, Ya Li, Hongjun He,\\ Zhongpei Zhang, \IEEEmembership{Member,~IEEE}, and Chadi Assi, \IEEEmembership{Fellow,~IEEE}
		
		%\thanks{This work was supported in part by the National Key Research and Development Program of China under Grant 2020YFB1806800. (\textit{Corresponding author:
				%Zhongpei~Zhang}.)
			%}
		
		\thanks{This work was supported in part by the Natural Science Foundation of Shenzhen City under Grant JCYJ20210324140002008;  in part by the Natural Science Foundation of Sichuan Province under Grant 2022NSFSC0489; and the joint project of China Mobile Research Institute \& X-NET.
			
			Yizhou Peng,	Songjie Yang, Wanting Lyu, 
			and Zhongpei Zhang are with the National Key Laboratory of Wireless Communications, University of Electronic Science and Technology of China, Chengdu 611731, China. 
			(e-mail: pengyizhou@std.uestc.edu.cn;
			yangsongjie@std.uestc.edu.cn;
			lyuwanting@yeah.net; 
			zhangzp@uestc.edu.cn).% <-this % stops a space
			
			Ya Li and Hongjun He are with the Future Research Laboratory, China Mobile Research Institute, Beijing, China
			(liyayjy@chinamobile.com;hehongjun@chinamobile.com)
			
			C. Assi is with Concordia
			University, Montreal, Quebec, H3G 1M8, Canada (email:assi@ciise.concordia.ca).}
	}
	\maketitle
	
	\begin{abstract}
		In this letter, a weighted minimum mean square error (WMMSE) empowered integrated sensing and communication (ISAC) method is investigated. One transmitting base station and one receiving wireless access point are considered to serve multiple users and a sensing target. Inspired by mutual information (MI), a unified framework to link sensing and communication is constructed, and communication MI and sensing MI rates are utilized as the performance metrics under the presence of clutters. In particular, we propose a novel MI-based WMMSE-ISAC method to maximize the weighted sensing and communication sum rate of this system. Such a maximization process is achieved by utilizing the classical method---WMMSE, aiming to better manage the effect of sensing clutters and the interference among users. Numerical results show the effectiveness of our proposed method, and the performance trade-off between sensing and communication is also validated.
	\end{abstract}
	\IEEEpeerreviewmaketitle
	\vspace{-1ex}
	\section{Introduction}
	\IEEEPARstart{R}{ecently}, research towards the construction of next-generation wireless networks(such as B5G and 6G) has increased dramatically. With the progression of innovative implementations such as vehicle to everything (V2X) and the Internet of Things (IoT), it is anticipated that the forthcoming wireless systems will play a crucial role in delivering various sensing services, encompassing a wide range of functionalities, e.g., target tracking and environmental monitoring. To achieve this, integrated sensing and communication (ISAC) provides a promising framework to seamlessly incorporate sensing functionalities within communication systems\cite{dual}.
	
	Under the ISAC framework, some very solid works have been made, such as \cite{ILA}-[4], and most of the existing works utilize achievable rates or signal-to-interference-plus-noise-ratio (SINR) as the communication metric [4]-[6]. Conversely, performance metrics for sensing vary. For instance, \cite{joint_bi} and \cite{time_bi-static} employ the probability of false alarm (PFA) to quantify sensing performance, while \cite{Per} applies signal-to-clutter-and-noise-ratio (SCNR), and [8]-[10] utilize the Cramer-Rao bound (CRB) to measure sensing performance. These miscellaneous metrics do not establish consistency with those for communication, often leading to complex optimization processes. Moreover, when building scenarios, most current works only focus on single-antenna users and ignore the influences of clutters, which fundamentally affect sensing performances in reality. These issues therefore motivate us to seek a novel ISAC framework with an easy, common, and reasonable sensing and communication (S\&C) measurement to address intricate scenarios when designing beamformers, such as multi-antenna users and sensing clutters. The concept of mutual information (MI) thus raises our attention. However, to our surprise, many current works about MI-based ISAC still apply inconsistent S\&C metrics and focus on simple scenarios\cite{MIISAC}\cite{JMI}, encouraging us to research further.
	
	According to \cite{mutual}, communication MI (CMI) has the well-known operational meaning of maximum achievable channel coding rate (i.e. CMI rate), which is directly related to SINR or SNR, while sensing MI (SMI) is similar to CMI both physically and mathematically. This feature motivates us to find a sensing metric consistent with CMI based on SMI. Inspired by \cite{Per}, which utilizes SCNR to measure its sensing performance, we thus set the SMI-based sensing rate (related to SCNR) as the sensing metric. It is hence possible for us to leverage some classical methods of communication-only frameworks in such ISAC cases to optimize its certain utility.

	In this letter, we propose a unified WMMSE-ISAC algorithm based on the MI framework\footnote{The code of the simulation of our proposed algorithm is available at https://github.com/ROCASSO/MI-based-WMMSE-ISAC-algorithm}. We consider a complex scenario with several multi-antenna users and a sensing target (ST), with sensing clutters presented to better simulate real-life situations\footnote{The objective of this letter is to probe whether the target exists in a known direction. For larger areas, an option is to design the transceiver to monitor over a range of directions}. A beamforming design problem is then formulated to maximize the weighted S\&C sum rate for sensing and communication, subject to a maximum power constraint. By utilizing the well-known weighted minimum mean square error (WMMSE) method, the non-convex problem is then transformed into a convex and derivable one. Therefore, the problem is solved through a simple iterative process. The effectiveness of this algorithm was finally verified by our numerical results.
	
	\emph{Notations}: Bold lower-case and upper-case letters denote column vectors and matrices respectively. Standard lower-case letters represent constants or variables. $(\cdot)^\HH$ represents the Hermitian operation, while $(\cdot)^{-1}$ and $\rm Tr(\cdot)$ stand for inversion and trace operation correspondingly. $\mathbb{C}^{n \times m}$ denotes the ${n \times m}$ complex space, and $\mathbb{E}[\cdot]$ is the expectation of a random variable.

	\begin{figure}[h]
		\centering
		\includegraphics[width=2.3in, height=2.3in]{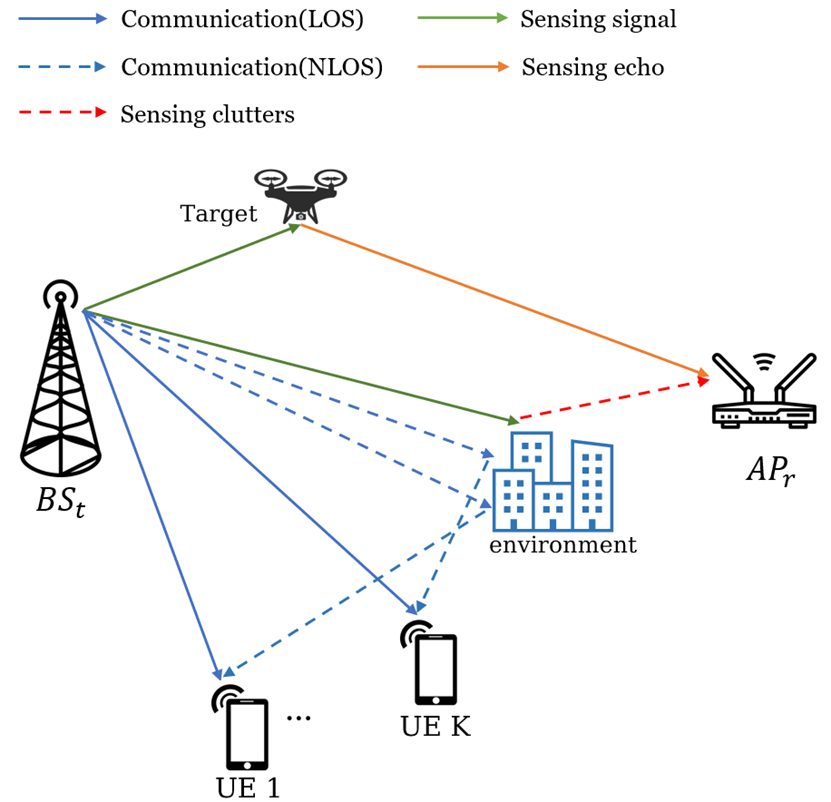}
		\caption{Illustration of MI-based WMMSE-ISAC system.}
		\label{sys1}
	\end{figure}
	
	\vspace{-2ex}
	\section{System Model}

	We consider a bi-static ISAC system, which consists of one transmitting base station ($\text{BS}_{t}$) and one receiving wireless access point ($\text{AP}_{r}$) with $K$ communication user equipment (UEs) and one ST. $\text{AP}_{r}$ and $\text{BS}_{t}$ are connected through high-capacity links and form the sensing network. In our proposed system, the main focus is on interference management and utility maximization for S\&C through transceiver design. In addition, sensing clutters that affect the sensing performance are considered. The transmitting and receiving BSs are equipped with $N_{t}$ and $N_{r}$ antennas respectively, and each UE is equipped with $R$ antennas. Without the loss of generality, one data stream is allocated for each antenna of a UE, and $S$ data streams are assumed for the ST. The set for all UEs is thus written as $\mathcal{I} = \left\{i | i \in {{1, \cdots, K}}\right\}$.
	Moreover, the channel matrix $\mathbf{H}$ for UEs is given by:
	\begin{equation}
		\begin{split}
			\mathbf{H}=\left[{\mathbf{H}_{1}, \cdots,\mathbf{H}_{K}} \right] \in \mathbb{C}^{R\times KN_t},
		\end{split}
	\end{equation}
	and we name the ST as sensing target $\tau.$ ${\mathbf{G}_{\tau}} \in \mathbb{C}^{N_r\times N_t}$ stands for the channel matrix for the ST. The transmitting beamformer matrix for both sensing and communication is given as:
	\begin{small}
		\begin{equation}
			\begin{split}
				\mathbf{V}=\left[ \underbrace{\mathbf{V}_{1}, \cdots,\mathbf{V}_{K}}_{\text{Communication}},\underbrace{\mathbf{V}_{{\tau}}}_{\text{Sensing}} \right] \in \mathbb{C}^{N_t\times (KR+S)}.
			\end{split}
		\end{equation}
	\end{small}

	Let $\mathbf{V}_{\tau}$ denotes the beamformer for sensing target $\tau$ and $\mathbf{V}_{k}$ for user ${k}$, and the corresponding signal is $\mathbf{s}_{\tau} \in \mathbb{C}^{S \times 1}$ and $\mathbf{s}_{k}$ $\in \mathbb{C}^{R \times 1}$, where $\mathbb{E}[\mathbf{s}_{k} \mathbf{s}_{k}^\HH] = \mathbb{E}[\mathbf{s}_{\tau} \mathbf{s}_{\tau}^\HH] = 1$. 
	
	\subsection{Communication Model} 
	In this letter, the well-known Saleh-Valenza (SV) model is utilized for channel construction \cite{eng}. The  signal transmitted to user ${k}$ can be written as: 
	\begin{small}
		\begin{equation}
			\begin{split}
				\mathbf{y}_{k}&=\underbrace{\mathbf{H}_{k} \mathbf{V}_{k}  \mathbf{s}_{k}}_{\text{expected signal}}+\underbrace{\sum_{i=1,i\neq k}^{K} \mathbf{H}_{k} \mathbf{V}_{i} \mathbf{s}_{i}}_{\text{multi-user interference}}+\underbrace{\mathbf{H}_{k} \mathbf{V}_{\tau} \mathbf{s}_{\tau}}_{\text{sensing interference}}+\mathbf{n}_{c},
			\end{split}
		\end{equation}
	\end{small}
	where $\mathbf{n}_{c} \in \mathbb{C}^{R\times1}$ is the additive white Gaussian noise with the distribution $\mathcal{N}(0,\sigma_{c}^{2})$. The channel $\mathbf{H}_{k}$ here is given by 
	\begin{equation}
		\begin{split}
			\mathbf{H}_{k}=\gamma \sum_{p=1}^{P} \beta_{k,p} \mathbf{a}_{R} \mathbf (\phi_{k,p}) \mathbf{a}_{t}^\HH \mathbf(\varphi_{k,p}),
		\end{split}
	\end{equation}
	where $\gamma = \sqrt{\frac{{N_{t}} {R}}{P}}$. $\mathbf{a}_{t}(\phi_{k,p})$ and $\mathbf{a}_{R}(\varphi_{k,p})$ are the steering vectors in $\text{BS}_{t}$---$\text{UE}$ channels with $\vert\vert\mathbf{a}_{t}(\phi_{k,p})\vert\vert ^2 = \vert\vert\mathbf{a}_{R}(\varphi_{k,p})\vert\vert ^2 = 1$. It is assumed that $P$ paths exist in the environment. $\phi_{k,p}$ and $\varphi_{k,p}$ denote the AOA and AOD at the $p^{th}$ path between the $\text{BS}_{t}$ and the $k^{th}$ UE respectively, and $\beta_{k,p} \sim \mathcal{N}(0,\sigma_{k,p}^2)$ is the path gain.
	\vspace{-0.3ex}
	\subsection{Sensing Model}
	The effective received signal to detect the sensing target at $\text{AP}_r$ is:
	\begin{footnotesize}
		\begin{equation}
			\begin{split}
				\mathbf{y}_{\tau}=\underbrace{\mathbf{G}_{\tau} \mathbf{V}_{\tau}  \mathbf{s}_{\tau}}_{\text{self}}+\underbrace{\sum_{i=1}^{K} \mathbf{G}_{\tau} \mathbf{V}_{i} \mathbf{s}_{i}}_{\text{user interference}}
				+\underbrace{\sum_{l=1}^{L}\mathbf{G}_{l}\mathbf{V}_{\tau}\mathbf{s}_{\tau}+\sum_{l=1}^{L}
					\sum_{j=1}^{K}\mathbf{G}_{l}\mathbf{V}_{j}\mathbf{s}_{j}}_{\text{sensing clutter}}+\mathbf{n}_{s},
			\end{split}
		\end{equation}
	\end{footnotesize}where  $\mathbf{n}_{s} $ is similar to $\mathbf{n}_{k}$, following the distribution $ \mathcal{N}(0,\sigma_{s}^2)$. The $\text{BS}_{t}-\text{ST}-\text{AP}_{r}$ channel $\mathbf{G}_{\tau}$ and the $\text{BS}_{t}-\text{clutter patch}-\text{AP}_{r}$ channel $\mathbf{G}_{l}$ are given by: 
	\begin{subequations}
		\begin{align}
			\mathbf{G}_{\tau} &= \beta_{\tau} \mathbf{a}_{R} \mathbf (\phi_{\tau}) \mathbf{a}_{t}^\HH \mathbf(\varphi_{\tau}),\\
			\mathbf{G}_{l} &= \beta_{l}\mathbf{a}_{R} \mathbf (\phi_{l}) \mathbf{a}_{t}^\HH \mathbf(\varphi_{l}).
		\end{align}
	\end{subequations}
	Here, $\phi_{\tau}$ and $\varphi_{\tau}$ represent the AOA and AOD of the  $\text{BS}_{t}-\text{ST}-\text{AP}_{r}$ channel for the ST, and $\phi_{l}$ and $\varphi_{l}$ stand for the AOA and AOD at the $l^{th}$ $\text{BS}_{t}-\text{clutter patch}-\text{AP}_{r}$ channel respectively, with $\vert\vert\mathbf{a}_{t}(\phi)\vert\vert ^2 = \vert\vert\mathbf{a}_{R}(\varphi)\vert\vert ^2 = 1$. Their channel gains are denoted as $\beta_{\tau}\sim\mathcal{N}(0,\sigma_{\tau}^2)$ and $\beta_{l}\sim\mathcal{N}(0,\sigma_{l}^2)$ respectively. $L$ clutter patches are assumed to exist under such a scenario.

	\vspace{-1ex}
	\section{Problem Formulation}
	We utilize SMI and CMI rates as the metrics for the entire ISAC system. The CMI rate is the maximum achievable channel coding rate, while the meaning of SMI varies\cite{mutual}. However, SMI is similar to CMI both physically and mathematically \cite{mutual}. According to \cite{Per}, it is well known that detection probability is an increasing function of SCNR. To resemble the CMI rate, we exploit the SMI-based sensing rate, namely the SMI rate, as our metric, which is $\log\det(\mathbf{I}+\text{SCNR})$. We consider receive beamforming strategy as follows:
	\begin{equation}
		\begin{split}
			\mathbf{\widetilde{s}}_{k} = \mathbf{B}_{k}^\HH \mathbf{s}_{k},
			\, \mathbf{\widetilde{s}}_{\tau} = \mathbf{B}_{\tau}^\HH \mathbf{s}_{\tau}.
		\end{split}
	\end{equation}
	With the similarity between S\&C metrics, the problem is formulated as a classical weighted sum-rate maximization one. The key of the problem is then set to optimize the beamformers to maximize the weighted S\&C sum-rate of this system while ensuring the power budget, which is $\sum_{k=1}^{K} {\rm Tr}(\mathbf{V}_{k} \mathbf{V}_{k}^\HH)+{\rm Tr}(\mathbf{V}_{\tau} \mathbf{V}_{\tau}^\HH) \leq {P}_{0}$.
	\vspace{1ex}
	
	The weighted S\&C sum-rate maximization problem is:
	\begin{small}
		\begin{equation}
			\begin{split}
				&\max_{\mathbf{V}}\sum_{k=1}^{K} \mathbf{\alpha}_{k} {R}_{k}+\mathbf{\alpha}_{\tau} {R}_{\tau}\\ 
				&s.t. \sum_{k=1}^{K} {\rm Tr}(\mathbf{V}_{k} \mathbf{V}_{k}^\HH)+{\rm Tr}(\mathbf{V}_{\tau} \mathbf{V}_{\tau}^\HH)\leq {P}_{0},
			\end{split}		
		\end{equation}
	\end{small}
	where  $\alpha_k$, $\alpha_{\tau} > 0 $ represent the weighting coefficients, while
	\begin{footnotesize}
		\begin{subequations}
			\begin{align}
				{R}_k &\triangleq \log\det \Big(\mathbf{I}+\mathbf{H}_{k} \mathbf{V}_{k} \mathbf{V}_{k}^\HH \mathbf{H}_{k}^\HH(\sum\limits_{i=1,i\neq k}^{K} \mathbf{H}_{k} \mathbf{V}_{i} \mathbf{V}_{i}^\HH \mathbf{H}_{k}^\HH
				+\mathbf{H}_{k} \mathbf{V}_{\tau} \mathbf{V}_{\tau}^\HH \mathbf{H}_{k}^\HH+\sigma_{k}^2 \mathbf{I})^{-1} \Big) ,\forall k \in \mathcal{I}\\
				{R}_{\tau} &\triangleq \log\det\Big(\mathbf{I}+\mathbf{G}_{\tau} \mathbf{V}_{\tau} \mathbf{V}_{\tau}^\HH \mathbf{G}_{\tau}^\HH (\sum\limits_{j=1}^{K} \mathbf{G}_{\tau} \mathbf{V}_{j} \mathbf{V}_{j}^\HH \mathbf{G}_{\tau}^\HH
				+ \sum\limits_{l=1}^{L} (\sum\limits_{n=1}^{K} \mathbf{G}_{l} \mathbf{V}_{n} \mathbf{V}_{n}^\HH \mathbf{G}_{l}^\HH + \mathbf{G}_{l} \mathbf{V}_{\tau} \mathbf{V}_{\tau}^\HH \mathbf{G}_{l}^\HH)+\sigma_{\tau}^2 \mathbf{I})^{-1}).  		 
			\end{align}
		\end{subequations}
	\end{footnotesize}
	$R_k$ and $R_{\tau}$ denote the CMI rate for the $k^{th}$ UE and the SMI rate for the ST respectively. When calculating $R_{\tau}$, the clutter is strong enough to ignore other communication signals (i.e., $\sum_{j=1}^{K} \mathbf{G}_{\tau} \mathbf{V}_{j} \mathbf{V}_{j}^\HH \mathbf{G}_{\tau}^\HH$) at the path $\mathbf{G}_{\tau}$. The SMI rate ${R}_{\tau}$ can thus be interpreted as the $\log\det(\mathbf{I}+\text{SCNR})$ form, which is directly related to SCNR.
	
	\vspace{1ex}
	Following \cite{5756489}, the receiving beamformer $\mathbf{B}_{k}$ and $\mathbf{B}_{\tau}$ can be easily determined as:
	\begin{footnotesize}
		\begin{equation}
			\begin{split}
				\mathbf{B}_{k}=	\mathbf{A}_{k}^{-1} \mathbf{H}_{k} \mathbf{V}_{k}\,\text{,}\,\, 
				\mathbf{B}_{\tau}=	\mathbf{A}_{\tau}^{-1} \mathbf{G}_{\tau} \mathbf{V}_{\tau}\,\text{.} \label{Bk}
			\end{split}
		\end{equation}
	\end{footnotesize}
	the corresponding MSE matrices are thus written as:
	\begin{footnotesize}
		\begin{subequations}
			\begin{align}
				\mathbf{E}_{k}=&\mathbf{I}- \mathbf{V}_{k}^\HH \mathbf{H}_{k}^\HH \mathbf{A}_{k}^{-1} \mathbf{H}_{k} \mathbf{V}_{k}\, \text{,}\\
				\mathbf{E}_{\tau}=& \mathbf{I}- \mathbf{V}_{\tau}^\HH \mathbf{G}_{\tau}^\HH \mathbf{A}_{\tau}^{-1} \mathbf{G}_{\tau} \mathbf{V}_{\tau}\, \text{,}
				\label{Bi}
			\end{align}
		\end{subequations}
	\end{footnotesize}
	where
	\begin{footnotesize}
		\begin{subequations}
			\begin{align}
				\mathbf{A}_{k}=& \sum\limits_{i=1}^{K} \mathbf{H}_{k} \mathbf{V}_{i} \mathbf{V}_{i}^\HH \mathbf{H}_{k}^\HH +\mathbf{H}_{k} \mathbf{V}_{\tau} \mathbf{V}_{\tau}^\HH \mathbf{H}_{k}^\HH+\sigma_{k}^{2}\mathbf{I}\text{,}\\
				\mathbf{A}_{\tau}=& \mathbf{G}_{\tau} \mathbf{V}_{\tau} \mathbf{V}_{\tau}^\HH \mathbf{G}_{\tau}^\HH + \sum\limits_{j=1}^{K}\mathbf{G}_{\tau} \mathbf{V}_{j} \mathbf{V}_{j}^\HH \mathbf{G}_{\tau}^\HH
				+\sum\limits_{l=1}^{L} \Big(\sum\limits_{n=1}^{K} \mathbf{G}_{l} \mathbf{V}_{n} \mathbf{V}_{n}^\HH \mathbf{G}_{l}^\HH
				+\mathbf{G}_{l} \mathbf{V}_{\tau} \mathbf{V}_{\tau}^\HH \mathbf{G}_{l}^\HH\Big)+\sigma_{\tau}^{2} \mathbf{I}\,
			\end{align}
		\end{subequations}
	\end{footnotesize}
	respectively denote the covariance matrix of all signals received at the receivers of UE $k$ and the ST. 
	\vspace{-1ex}
	\section{Proposed Solution}
	To handle the non-convex weighted S\&C sum-rate maximization problem, we utilize the classical WMMSE method \cite{5756489} to transform it into a convex one, which is given by: 
	\begin{small}
		\begin{equation}
			\begin{split}
				\min_{\mathbf{V,B,W}} &\bigg[\sum_{k=1}^{K} \mathbb{\alpha}_{k} \big({\rm Tr}(\mathbf{W}_{k} \mathbf{E}_{k})-\log\det(\mathbf{W}_{k})\big)
				+ \mathbb{\alpha}_{\tau} \big({\rm Tr}(\mathbf{W}_{\tau} \mathbf{E}_{\tau})-\log\det(\mathbf{W}_{\tau})\big)\bigg] \quad \\ 
				s.t. &\quad \sum_{k=1}^{K} {\rm Tr}(\mathbf{V}_{k} \mathbf{V}_{k}^\HH)+ {\rm Tr}(\mathbf{V}_{\tau} \mathbf{V}_{\tau}^\HH)\leq {P}_{0}. \label{problem1}
			\end{split}		
		\end{equation}
	\end{small}
	
	The problem can then be fixed in the space of $\left\{\mathbf{B}, \mathbf{V}, \mathbf{W}\right\}$. Updating one variable with others remaining fixed makes this problem easy to solve. By applying the Lagrangian multiplier method, the problem can be rewritten as an unconstrained one, which is given by:
	\begin{small}
		\begin{equation}
			\begin{split}
				\min_{\mathbf{V,B,W}} &\sum_{k=1}^{K} \mathbb{\alpha}_{k} ({\rm Tr}(\mathbf{W}_{k} \mathbf{E}_{k})-\log\det(\mathbf{W}_{k}))
				+ \mathbb{\alpha}_{\tau} ({\rm Tr}(\mathbf{W}_{\tau} \mathbf{E}_{\tau})-\log\det(\mathbf{W}_{\tau}))\\
				&+ \lambda (\sum_{k=1}^{K}{\rm Tr}(\mathbf{V}_{k} \mathbf{V}_{k}^\HH)+ {\rm Tr}(\mathbf{V}_{\tau} \mathbf{V}_{\tau}^\HH)-{P}_{0}).
			\end{split}		
		\end{equation}
	\end{small}
	
	We name its objective function as $f$. Due to its convex feature for each of the optimization variables \cite{5756489}, the block coordinate descent method is applied. Sequentially, after fixing two of the optimization variables, the left one can be solved and updated simultaneously for all users and sensing targets. Inspired by \cite{5756489}, the update of the weight matrix $\mathbf{W}_{k}$ or $\mathbf{W}_{\tau}$ is in closed form, which is the inverse of the corresponding MSE matrix, making it easy to update. The solution of receiving beamformer $\mathbf{B}_{i}$ comes from (11), and the updating processes for all $\mathbf{V}_{i}$ are independent from each other. In this case, we first consider the ${k}^{th}$ beamformer which belongs to one UE. Following the methods provided by \cite{weighted}, fundamental components of $\nabla _{\mathbf{V}_{k}} f$ can be derived as:
	
	\begin{footnotesize}
		\begin{equation}
			\begin{split}
				\nabla_{\mathbf{V}_{k}} {\rm Tr}(\mathbf{W}_{k} \mathbf{E}_{k})=2\alpha_{k} \mathbf{H}_{k}^\HH \mathbf{B}_{k} \mathbf{W}_{k} \mathbf{B}_{k}^\HH \mathbf{H}_{k} \mathbf{V}_{k}
				-2\mathbb{\alpha}_{k}
				\mathbf{H}_{k}^\HH \mathbf{B}_{k} \mathbf{W}_{k}, \label{self}    
			\end{split}		
		\end{equation}
	\end{footnotesize}
	\vspace{-3ex}
	\begin{footnotesize}
		\begin{flalign}
			&\qquad\qquad\qquad\qquad\qquad\qquad\nabla_{\mathbf{V}_{k}} {{\rm Tr}(\mathbf{W}_{i} \mathbf{E}_{i})} =& \notag
		\end{flalign}
	\end{footnotesize}
	
	\vspace{-3ex}
	\begin{footnotesize}
		\begin{equation}
			\begin{cases}
				2\alpha_{i} \mathbf{H}_{i}^\HH \mathbf{B}_{i} \mathbf{W}_{i} \mathbf{B}_{i}^\HH \mathbf{H}_{i} \mathbf{V}_{k}, 
				\forall i = 1,\cdots,K, i\neq k, \\
				2\alpha_{i} \mathbf{G}_{i,s}^\HH \mathbf{B}_{i} \mathbf{W}_{i} \mathbf{B}_{i}^\HH \mathbf{G}_{i,s} \mathbf{V}_{k}+2\alpha_{i} \sum\limits_{l=1}^{L}\mathbf{G}_{l}^\HH \mathbf{B}_{i} \mathbf{W}_{i} \mathbf{B}_{i}^\HH \mathbf{G}_{l} \mathbf{V}_{k},
				
				i=\tau. \label{others}
			\end{cases}		
		\end{equation}
	\end{footnotesize}		
	(\ref{self}) is the key component relevant to the UE itself, while (\ref{others}) is the key component about other users and the sensing target. With regard to the ST, the key components are fairly similar.
	
	Through combining these components altogether properly, the derivatives of $f$ with respect to the   beamformer for the $k^{th}$ UE and the ST $\tau$ can be directly obtained. Via setting the derivatives to zero, the optimized beamformers $\mathbf{V}_{k}^{opt}$ ($\forall k\in \mathcal{I}$) and $\mathbf{V}_{\tau}^{opt}$ are given by:   
	
	\begin{footnotesize}
		\begin{subequations}
			\begin{align}
				\mathbf{V}_{k}^{opt}=&\Big(\sum\limits_{i=1}^{K}\alpha_{i} \mathbf{H}_{i}^\HH \mathbf{B}_{i} \mathbf{W}_{i} \mathbf{B}_{i}^\HH \mathbf{H}_{i}+ \alpha_{\tau} \mathbf{G}_{\tau}^\HH \mathbf{B}_{\tau} \mathbf{W}_{\tau} \mathbf{B}_{\tau}^\HH \mathbf{G}_{\tau} 
				+\alpha_{\tau}\sum\limits_{l=1}^{L}\mathbf{G}_{l}^\HH \mathbf{B}_{\tau} \mathbf{W}_{\tau} \mathbf{B}_{\tau}^\HH \mathbf{G}_{l}+\lambda \mathbf{I}\Big)^{-1}\alpha_{k} \mathbf{H}_{k}^\HH \mathbf{B}_{k} \mathbf{W}_{k}\label{Vopt},\\
				\mathbf{V}_{\tau}^{opt}=&\Big(\sum\limits_{i=1}^{K}\alpha_{i} \mathbf{H}_{i}^\HH \mathbf{B}_{i} \mathbf{W}_{i} \mathbf{B}_{i}^\HH \mathbf{H}_{i} + \alpha_{\tau} \mathbf{G}_{\tau}^\HH \mathbf{B}_{\tau} \mathbf{W}_{\tau} \mathbf{B}_{\tau}^\HH \mathbf{G}_{\tau}
				+\alpha_{\tau}\sum\limits_{l=1}^{L}\mathbf{G}_{l}^\HH \mathbf{B}_{\tau} \mathbf{W}_{\tau} \mathbf{B}_{\tau}^\HH \mathbf{G}_{l}+\lambda \mathbf{I}\Big)^{-1}\alpha_{\tau} \mathbf{G}_{\tau}^\HH \mathbf{B}_{\tau} \mathbf{W}_{\tau}\label{Voptt}, 
			\end{align}		
		\end{subequations}
	\end{footnotesize}
	
	Notably, $\lambda\geq{0}$ is the parameter that ensures the power constraint. Since it can be shown that $\sum_{k=1}^{K}{\rm Tr}(\mathbf{V}_{k}^{opt}(\mathbf{V}_{k}^{opt}))+{\rm Tr}(\mathbf{V}_{\tau}^{opt} (\mathbf{V}_{\tau}^{opt}))$ is a decreasing function with respect to $\lambda$ \cite{5756489}, its value can be easily determined through the bisection approach. It turns out that the optimized beamformers for sensing and communication are actually identical. This results from the mutual-information framework and the S\&C metrics we follow, bridging communication and sensing altogether to make the system more integrative. The WMMSE-ISAC algorithm is provided as Algorithm 1, and \cite[Theorem 2]{5756489} guarantees the algorithm to converge to a stationary point of (\ref{problem1}). Let $\iota = K+1$ denote the total number of UEs and the ST. The complexity of each iteration is $\mathcal{O}(\iota N_{t}^{3} + K R^{3} + N_{r}^{3})$ for one user's or the ST's beamformer.
	\begin{footnotesize}
		\begin{algorithm}[h]
			\caption{Proposed WMMSE-ISAC algorithm to solve problem (\ref{problem1}) }   
			\textbf{Input} Initialize $\mathbf{V}$ to meet  
			the power constraint of (\ref{problem1}) \\
			\quad\textbf{Repeat}
			\begin{enumerate} 
				\item Compute all $\mathbf{B}_{k}$,$\mathbf{B}  _{\tau}$ based on (10) 
				\item Calculate all $\mathbf{E}_{k}$,$\mathbf{E}  _{\tau}$ based on (11)
				\item Apply $\mathbf{W}_{k} \leftarrow \mathbf{E}_{k}^{-1},$ $\mathbf{W}_{\tau} \leftarrow \mathbf{E}_{\tau}^{-1}\, \text{,}\forall k\in \mathcal{I}$
				\item Update $\mathbf{V}_{k}$ and $\mathbf{V}_{\tau}$ according to (17), $\forall k\in \mathcal{I}$
			\end{enumerate} 
			\textbf{Until} Convergence criterion is met.\\
			\textbf{Output:} The optimal solution $\mathbf{V}_{k}$ and $\mathbf{V}_{\tau}$, , $\forall k\in \mathcal{I}$
			\label{ALG}
		\end{algorithm}
	\end{footnotesize}
	
	To see the optimization effects the weighting coefficients posing to the system, we set $\sum_{k=1}^{K}\alpha_{k} = \omega_{c}$ and $\alpha_{\tau}=\omega_{\tau}$ with $\omega_{\tau}\in [0,1)$ and $\omega_{c}=1-\omega_{\tau}$. When $\omega_{\tau}$ varies, the optimization performance of the WMMSE-ISAC system also needs to change to have a better response towards particular requirements.
	\vspace{-0.5cm}
	\section{Simulation Results}   
	In this section, numerical results are presented under a sub-6G system operating at $3.3$GHz to verify the effectiveness of our WMMSE-ISAC algorithm. All the array elements spacing is half of the wavelength. We assume that both $\text{BS} _{t}$ and $\text{AP}_{r}$ are equipped with ULAs, and $ N_{t} = 16$ while $N_{r}=4$, serving $K=3$ users. Only $S = 1$ data stream is allocated to sensing as it is theoretically sufficient for one sensing target. The direction of all UEs and the ST is randomly generated within the range of $[-\frac{\pi}{2},\frac{\pi}{2}]$. Each UE is assumed to have a ULA with $R = 4$. $L=3$ sensing clutter patches are set, while $P = 10$ paths are assumed in the environment. For clarity and simplicity of controlling the signal-to-noise-ratio (SNR) of the overall system, $\sigma_c^2 = \sigma_{s}^2$ are set to be $30$ dBm. We also set $\sigma_{k,p}^2 = 30$ dBm for the line-of-sight (LOS) path and $\sigma_{k,p}^2 = 20$ dBm for the non-line-of-sight (NLOS) paths. As for the sensing channel, $\sigma_{\tau}^2$ is set to be $30$ dBm.
	$\frac{1}{L}\sum_{l=1}^{L}\sigma_{l}^2$ is set to be equal to $\sigma_{\tau}^2$ to better justify the effectiveness of our algorithm under evident clutter interference. Additionally, the transmitting power limit is set based on different transmitting SNRs (i.e., $P_0 = 10^{\frac{\text{SNR}}{10}}$).
	
	For our proposed WMMSE-ISAC algorithm, the tolerance for the weighted S\&C sum rate between two iterations is 1$e^{-3}$, and the maximum number of each iteration process is 50.

	\vspace{-2ex}
	\subsection{System performance and S\&C trade-off}
\begin{figure}[h]
		\centering
		\includegraphics[width=3.0in,height=2.3in]{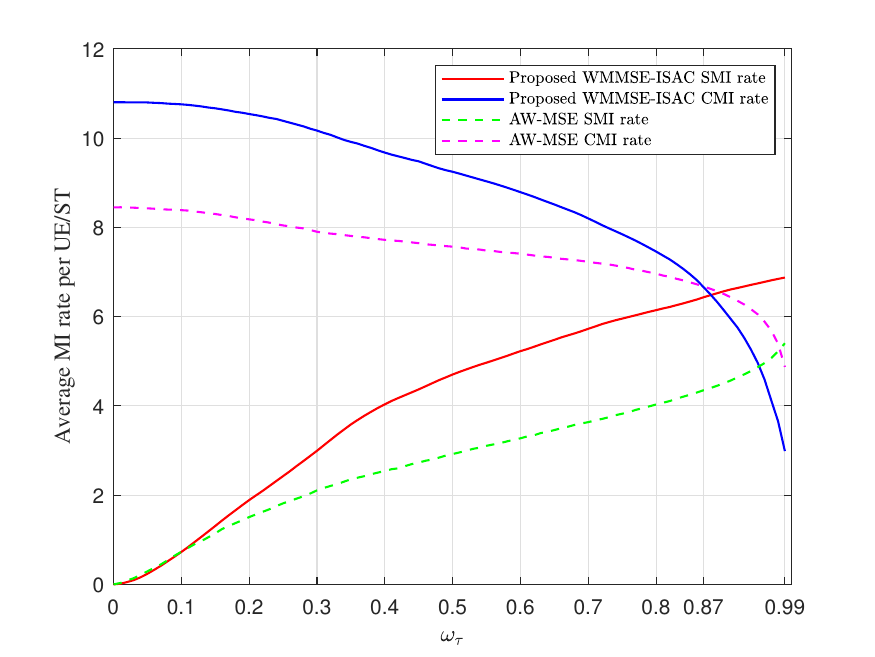}
		\caption{Average MI rate performance per UE/ST}
		\label{pic_1}
	\end{figure}

	Most existing conventional ISAC works only focus on single antenna users without receiving beamforming, making it hard for us to find a proper method to compare with. However, under our unified framework, classical methods for communication problems can be utilized in ISAC cases. We thus consider the classical Adaptively Weighted MSE (AW-MSE) transceiver design method\cite{MMSE} as our baseline. With the weighting coefficient $\omega_{\tau}$ ranging from 0.00 to 0.99, the average S\&C MI rate per UE/ST at $\text{SNR}=25$ dB is illustrated in Fig. \ref{pic_1}. For each curve, 500 Monte-Carlo experiments are performed. 
	
	According to Fig. \ref{pic_1}, as $\omega_{\tau}$ alters, both the CMI and SMI rates of the WMMSE-ISAC algorithm depict more apparent variations than those of the AW-MSE algorithm, indicating better sensitivity and flexibility towards the changes of the expected optimization preference. Such a feature results in a scarcely lower SMI rate and a much higher CMI rate when the optimization process is extremely communication-centric, and it also causes a lower CMI rate and a significantly better sensing performance when expecting a highly sensing-centric optimization ($\omega_{\tau}>0.87$).
	
\begin{figure}[h]
		\centering
		\includegraphics[width=3.0in,height=2.3in]{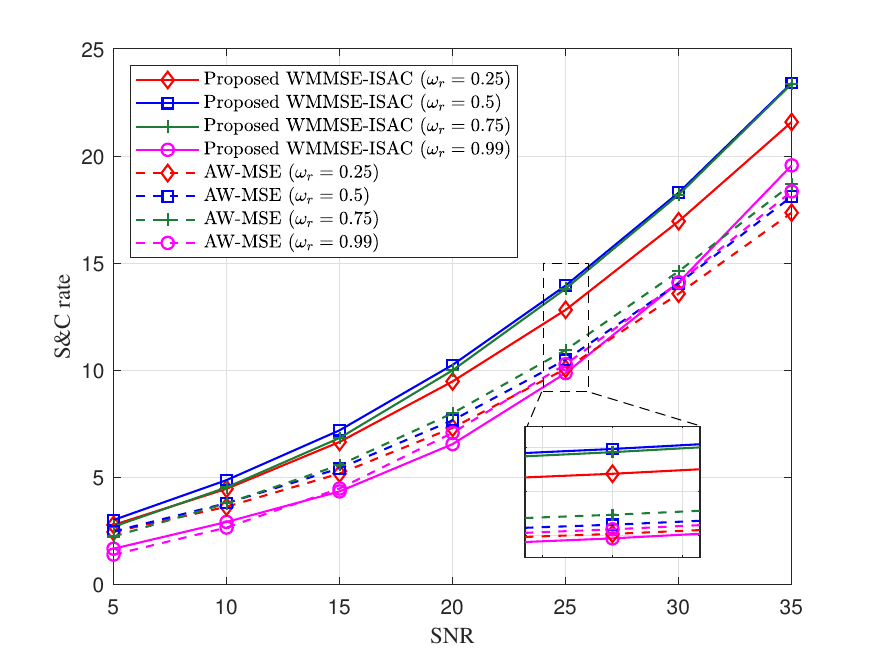}
		\caption{S\&C rate with different $\text{SNRs}$}
		\label{pic_2}
	\end{figure}

	As illustrated in Fig. \ref{pic_1}, the peak values of sensing and communication of our WMMSE-ISAC algorithm are also significantly higher than those of the AW-MSE method. With most of the curve of the SMI rate of our algorithm staying significantly higher than that of the baseline, we can conclude that our WMMSE-ISAC algorithm generally performs  better at sensing than the AW-MSE algorithm. With regard to communication performance, Fig. \ref{pic_1} clearly illustrates the superiority of our proposed algorithm when communication-centric optimization processes are expected ($\omega_{\tau}<0.5$). However, the CMI rate of our algorithm is slightly lower than that of our baseline when highly sensing-centric optimization is needed ($0.87<\omega_{\tau}<0.99$). In this case, the metric that much more matters is the SMI rate, which is significantly higher than that of our baseline. These phenomena above also show the sensitivity and successful S\&C performance trade-off of our algorithm towards the variation of the expected optimization preference, which are further justified in Fig. \ref{pic_2} and Fig. \ref{pic_3}.

	% We can thus conclude that the WMMSE-ISAC algorithm is more sensitive to the variation in the weighting coefficients, providing the system enough flexibility for different S\&C preferences. Such adaptability is further illustrated in Fig. \ref{pic_2}.
	
	Fig. \ref{pic_2} illustrates the average achievable rate per user plus the SMI rate of the ST (i.e., S\&C rate) under 4 situations with different expected optimization preferences. It is clear that the S\&C rate of our proposed WMMSE-ISAC algorithm outperforms those of the baseline in most cases. However, the S\&C rates of our algorithm when $\omega_{\tau} = 0.99$ are slightly lower than that of the AW-MSE method at some SNRs. Such a phenomenon is also due to the sensitivity and flexibility of our algorithm concluded from Fig.\ref{pic_1}. When $\omega_{\tau} = 0.99$, the CMI rate of our algorithm is lower than that of the AW-MSE method, while its SMI rate, the only metric that matters under such an extreme sensing-centric optimization scenario, is significantly higher, making the system better concentrated on sensing targets. However, with the AW-MSE algorithm, the S\&C rate performances at the given $\omega_{\tau}$s are quite close, indicating a not successful trade-off for the baseline algorithm. The overall performance trade-off will be further verified in Fig. \ref{pic_3}.
	
	\subsection{Overall Performance Trade-off}
	
\begin{figure}[h]
		\centering
		\includegraphics[width=3.0in,height=2.3in]{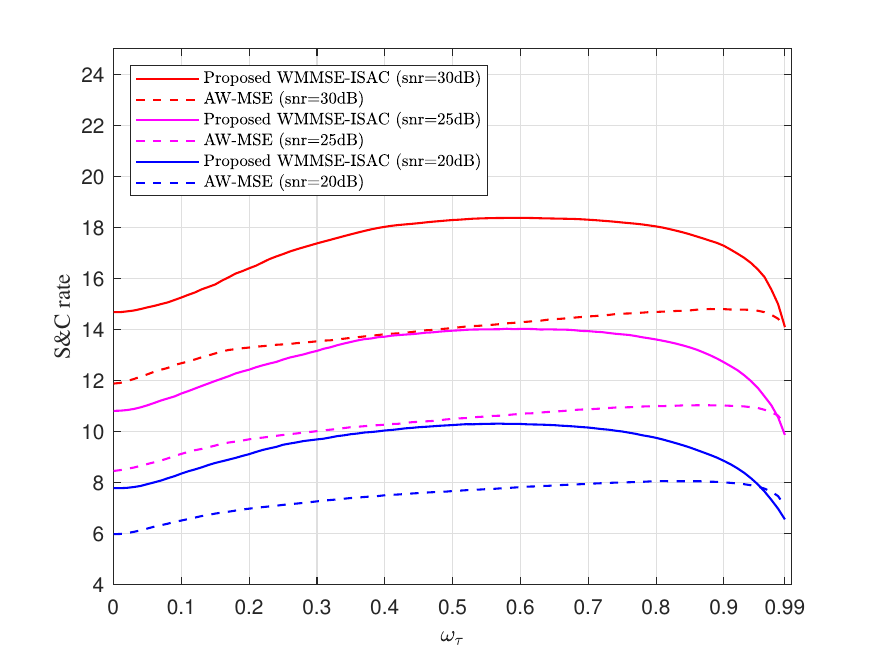}
		
		\caption{illustration of the S\&C rate variation with $\omega_{\tau}$ under different $\text{SNRs}$}
		\label{pic_3}
	\end{figure}

	Fig. \ref{pic_3} shows the S\&C rate variation with the value of $\omega_{\tau}$ at different SNRs. With our proposed WMMSE-ISAC algorithm, the peak values of the S\&C rate are significantly higher than the baseline. Moreover, it is also demonstrated that the system achieves a successful overall performance trade-off with our WMMSE-ISAC algorithm. For instance, the S\&C rate when $\omega_{\tau} = 0.25, 0.5$, and $0.75$ are substantially larger than those under extreme optimization preferences ($\omega_{\tau} = 0$ or $0.99$), indicating a much better performance when being utilized in general ISAC cases than extreme sensing-centric/communication-centric cases. Conversely, with the baseline algorithm, the S\&C rate is so insensitive to the variation of $\omega_{\tau}$ that it only shows slight elevation with $\omega_{\tau}$ increasing. Additionally, Fig. \ref{pic_1} shows even when the optimization focus is extreme, our algorithm still performs much better than the AW-MSE algorithm in terms of the corresponding sensing or communication metric. 
	
	\section{conclusions}
	In this letter, a unified WMMSE-ISAC framework has been proposed. Under the scenario of several multi-antenna users, one sensing target with sensing clutters, a joint S\&C beamforming design problem was formulated to maximize the weighted S\&C sum rate, which was tackled by the proposed algorithm. It turned out that our proposed algorithm was competent to effectively optimize the overall performance of the ISAC system under different optimization requirements. Generally speaking, our algorithm has the potential to provide high-quality sensing and communication services in complex scenarios.

	\vspace{-1.3ex}

\end{document}